# Accuracy and robust early detection of short-circuit faults in single-cell lithium battery

Chengzhong Zhang,[1,2] Hongyu Zhao,[2] and Wenjie Zhang,[1]

[1]College of Electrical and Power Engineering, Taiyuan University of Technology, Taiyuan 030024, PR China;
[2]University of Chinese Academy of Sciences, Shijingshan District, Beijing 100049, China

**CONTEXT & SCALE**

Lithium-ion batteries are widely used in electronics, electric vehicles, and renewable energy storage, playing a key role in green technology and a low-carbon economy. With the applications expansion, concerns about their safety, particularly thermal runaway, have gained widespread attention. Internal short circuits are the essential causes of thermal runaway in lithium-ion batteries, which can subsequently trigger safety incidents of combustion and explosion. Early detection of internal short-circuit failures is a challenging issue in this field due to their subtle fault characteristics. This paper presents a novel detection method in single cell, in which faults can be accurately and quickly identified. The details of this method listed in the text.

**SUMMARY**

Effective early-stage detection of internal short circuit in lithium-ion batteries is crucial to preventing thermal runaway. This report proposes an effective approach to address this challenging issue, in which the current change, state of charge and resistance are considered simultaneously to depict the voltage differential envelope curve. The envelope naturally utilizes the inherent physical information of the battery and accounts for error interference, providing a high-precision range for battery voltage fluctuations under any operating conditions. This study validates the algorithm using data from 10 fault intervals under dynamic operating condition. The results demonstrate that the algorithm achieves 100% accuracy and responds rapidly, enabling timely detection of early-stage internal short circuit faults in batteries. Compared to signal processing-based and neural network methods, the proposed approach offers significant advantages in both accuracy and practicality, making it highly relevant for the safe application and widespread adoption of lithium-ion batteries.

**INTRODUCTION**

Lithium-ion batteries (LIBs), due to their advantages of high energy density and power density, have been extensively applied in various fields such as electric vehicles, energy storage systems, and consumer electronics[1,2]. However, as the application continues to expand and the energy density further enhances, their safety issues have garnered increasing attention, particularly about the thermal runaway (TR)[3]. TR can release a substantial amount of heat in a short period, often accompanied by combustion and explosion, which presents significant challenges to the safe use and development of LIBs[4]. Feng et al[5]. has shown that internal short circuit faults (ISC) are the fundamental cause of TR in LIBs, although the triggers of TR can manifest in three distinct forms—mechanical abuse (e.g., mechanical deformation), thermal abuse (overheating), and electrical abuse (overcharging, external short circuits, etc.). The occurrence of ISC is random, and such faults can develop from the early to the late stages. In the later stages, ISC faults are typically accompanied by significant heat generation and changes in electrical characteristics[6] (voltage drop or deviation). Once the ISC reaches the late stage, it generally triggers TR immediately, making little practical significance, even if it can be detected. Therefore, early detection of ISC faults is an effective approach to preventing TR and remains a major focus and research hotspot in this field. However, early ISC faults exhibit extreme subtle changes in thermal and electrical characteristics, making their effective identification in the early stages a significant and almost unsolved challenge. Therefore, developing effective methods for early ISC faults detection is crucial for preventing TR and an urgent demand in this field.

ISC faults can be categorized into two major types: gradual-type (sort) and abrupt-type (hard). The former can be roughly divided into three levels—early, middle, and late stages—based on the heat generation power and equivalent short-circuit resistance. There is currently no clear standard for classifying abrupt ISC faults. Zheng et al[7,8]. demonstrated that early sort ISC faults could be identified on a battery pack by comparing the leakage characteristics of faulty battery with that of normal cells. Presently, the core idea of most ISC faults identification algorithms is to utilize the contrast information between faulty and normal batteries (e.g., state of charge (SOC) differences, voltage differences, and model parameter differences), as reviewed in Ref[9] and Ref[10]. In addition, the typical feature of ISC is a sharp voltage drop and recovery. The results presented in references[11,12] indicate that tens of millivolts fluctuations of voltage can occur in a hard-type ISC of battery. According to heat generation power and the degree of voltage change, this is considered an early ISC fault. Small voltage jumps are caused by the melting of the ISC loop. Since the structure near the short-circuit point is damaged, the short circuit is temporarily interrupted, but the fault tends to expand and evolve nearby. Therefore, in some cases, early-stage ISC voltage fluctuations occur even before the sustained leakage current characteristics, and identifying them at this moment would be more beneficial for the safe application of the LIBs.

To address the research gap in this field, this paper proposes an excellent algorithm for hard-type ISC faults identifying. The algorithm effectively utilizes the physical properties of the battery, enabling the early detection of ISC faults under any operating conditions. Compared to methods based on signal processing (such as signal decomposition, signal prediction, and signal reconstruction), information entropy theory, and deep learning approaches, this algorithm offers significant advantages. Moreover, it can operate effectively on individual cells without relying on comparative information from the module, ensuring excellent real-time performance and practicality. This is highly beneficial for enhancing the safety of lithium battery applications.

## RESULTS AND DISCUSSION

### Current processing and differential voltage calculation at normal state

The differential envelope curve of the battery voltage was defined to provide a reasonable variation range for normal battery under any operating conditions. Abnormal voltage fluctuations during the constant current and resting phases can be easily identified. Therefore, the application of the method proposed in this study during these two conditions is not presented. The dynamic stress test (DST) is utilized as the discharging schedule. Figure 1 shows the concrete current detail information and the current differential values over DST dynamic operation condition. As shown in Figure 1(a), the rounded value of the ratio of actual current to 0.05C (C denotes rate) is adopted for the envelope curve depict, which can provide a rational elastic variation space for the envelope and suppress the influence from noise. The differential values of current rounded values are demonstrated in Figure 1(c). Figure 1(b) and Figure 1(d) are partial enlargements of that of Figure 1(a) and Figure 1(c), respectively.

Figure 2 illustrates the voltage differential and the voltage differential envelope during the entire discharging process under the DST condition for a normal battery (excluding fault current pulses), in which blue solid line represents the actual voltage differential values, while the red dashed line denotes the envelope. Figure 2(b) and 2(c) are two local magnifications that provide a clearer view of the voltage differential and differential envelope in different SOC intervals, due to the nonlinearity between the internal resistance and SOC of lithium batteries. It can be clearly observed from Figure 2(b) and 2(c) that the envelope closely follows the actual voltage variations. At points of voltage sudden change, the voltage differential values remain within the bounds of the envelope, which indicates that the method proposed in this study (described in detail in the Methods section) is both reasonable and effective. Figure 2(d) presents the difference between the envelope and the voltage differential values. It is evident that their difference is influenced by both the current and the SOC. Significant changes in current lead to larger voltage fluctuations, and the amplitude of voltage fluctuations also increases in the low SOC range. This is attributed to the increase in internal resistance of the battery at lower SOC range. Under the DST condition, the difference between the battery voltage differential and the envelope generally remains within 0.01V, with the difference slightly increasing in the very low SOC range (<10%), but still staying around 0.012V. Figure 2(e) is a local magnification of Figure 2(d), which allows for a clearer observation of the relationship between the envelope and the actual voltage differential values.

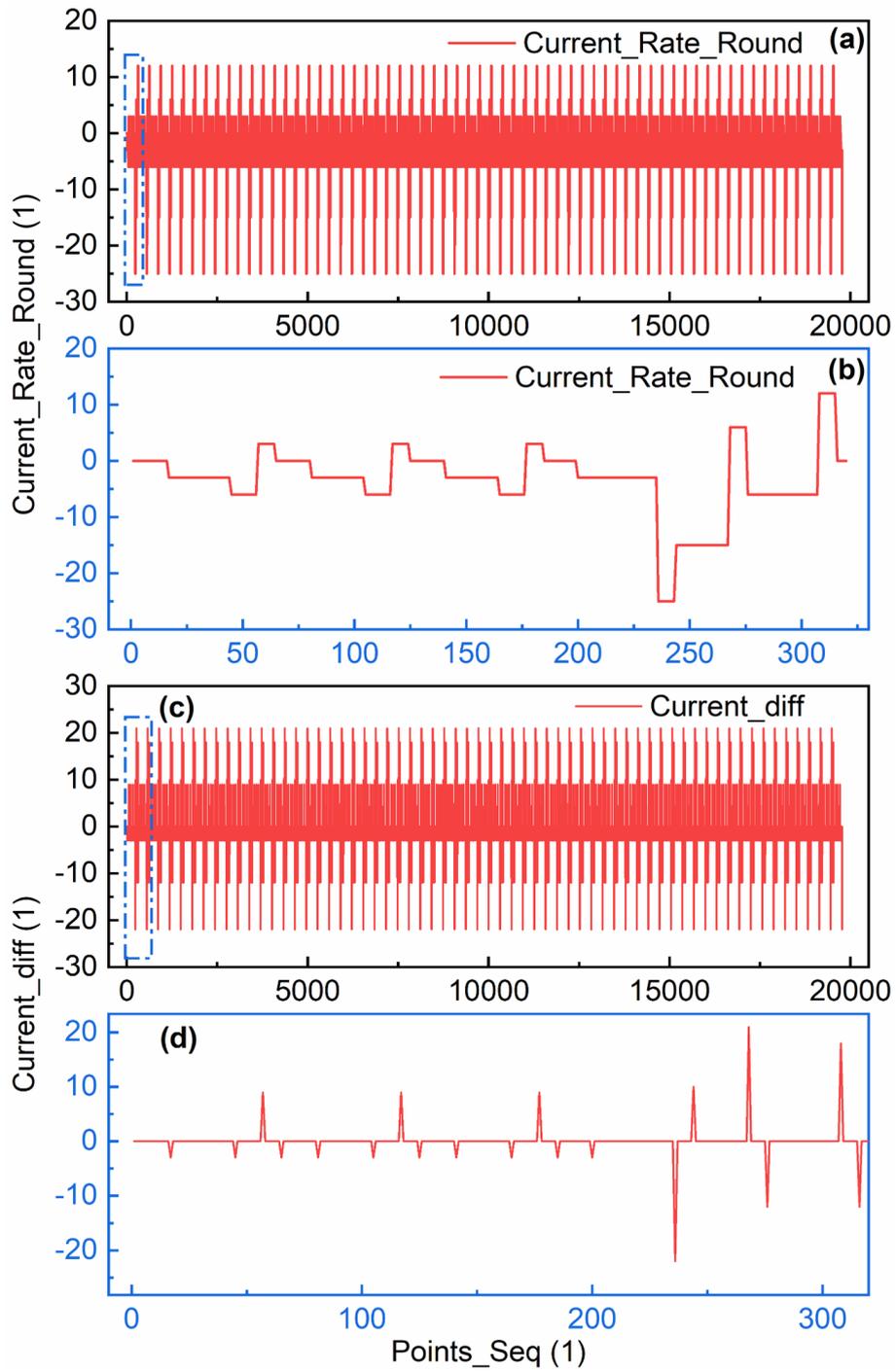

Figure 1. DST current schedule and current differential
(a) Rounded values of actual current to 0.05C
(b) Enlarged part of (a)
(c) Current differential values after rounding
(d) Enlarged part of (c)

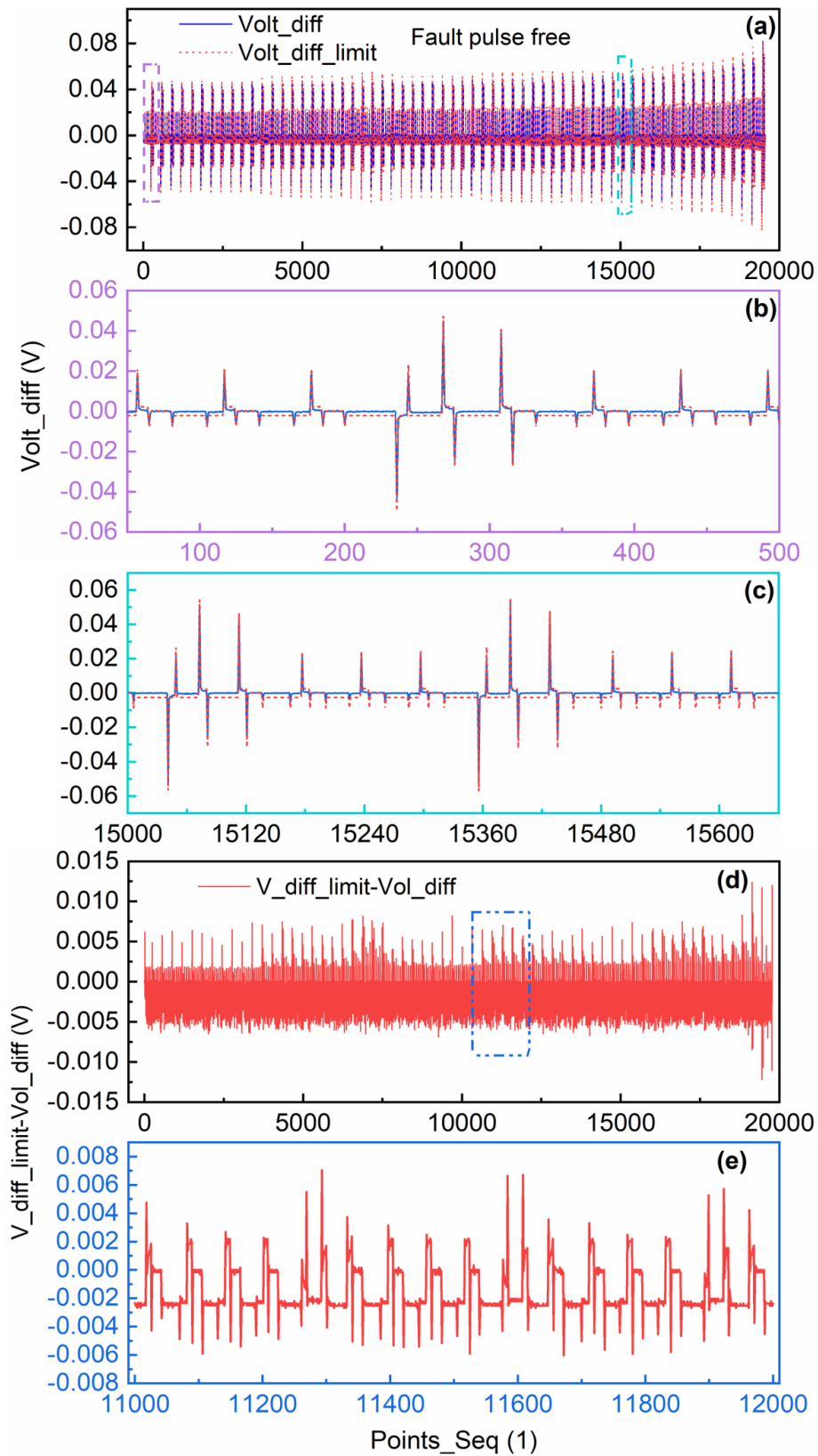

Figure 2. Voltage differential information and differential envelope curve of normal cell
(a) Values of voltage differential and envelope.
(b) The enlarged part over 50-500s of (a).
(c) The enlarged part over 15000-15660 of (a).

(d) The difference between the voltage differential and envelope.

(e) The enlarged part of (d).

**Information of faults and calculation results**

Figure 3(a) and 3(b) illustrate the distribution of fault pulse currents and the corresponding voltage information. In this study, ISC faults were simulated during the DST discharging period by connecting resistors of varying values. A total of 10 small fault intervals were set, each lasting approximately 30 seconds. From Figure 3(a), it can be seen that during fault intervals of larger current pulses, a noticeable voltage drop is observed. However, these faults belong to contextual anomaly pattern, and the current conditions used in the experiment are applied in a periodic manner. As a result, some fault intervals can be identified. On the other hand, when the fault pulse is smaller, as seen in the first and fifth fault intervals, the voltage drop is minimal and shows no significant deviation from surrounding sample points, making it almost impossible to identify under dynamic conditions. Furthermore, in real-world operational conditions, current fluctuations are far more complex than those in the laboratory settings, and they typically lack periodicity. Therefore, early detection of ISC faults is extremely challenging, especially at the scale of individual cells, where such identification is currently unachievable.

Figure 3(c) presents the actual voltage differential values and the voltage differential envelope calculated based on the battery's internal resistance and current differential, depicted by the blue solid line and red dashed line, respectively. It can be observed that, in the 10 fault intervals, the actual voltage differential consistently escapes from the envelope. To provide a clearer view of the voltage differential and envelope in the fault intervals, Figure 3(d) and 3(e) show magnified views of the first and fifth fault intervals, respectively. From the corresponding figures, an abnormal voltage jump pairs can be identified within respective fault interval. The first abnormal jump is a voltage drop, indicating a short circuit, while the second abnormal point exhibits a voltage increase, signaling that the short circuit path has been disconnected. This pattern of abnormal voltage variation corresponds well with the characteristic behavior of hard-type ISC faults. The interval between these two abnormal points corresponds to the fault duration. Therefore, the method proposed in this paper not only enables effective identification of hard-type ISC faults but also allows for precise fault localization. Moreover, the abnormal voltage variation manifests as paired anomalies, which leverages the inherent features of hard-type ISC faults to exclude isolated, incidental voltage jumps. This enhances the robustness of the algorithm.

**Faults detection results**

Figure 4 shows the difference between the voltage differential envelope and the actual voltage differential values over the entire DST discharge condition, including the fault intervals. The fault intervals are highlighted with red lines to enhance the visual effect. As previously analyzed, the main characteristic of hard-type ISC faults is the presence of a pair of abnormal voltage jump points in the voltage differential. Therefore, the curve in Figure 4 should exhibit this behavior within the fault intervals. From the results shown, it is evident that this pattern is present in all 10 fault intervals, which indicates that the method proposed in this paper can effectively identify ISC faults of varying degrees across all 10 fault scenarios, with an accuracy rate of 100%.

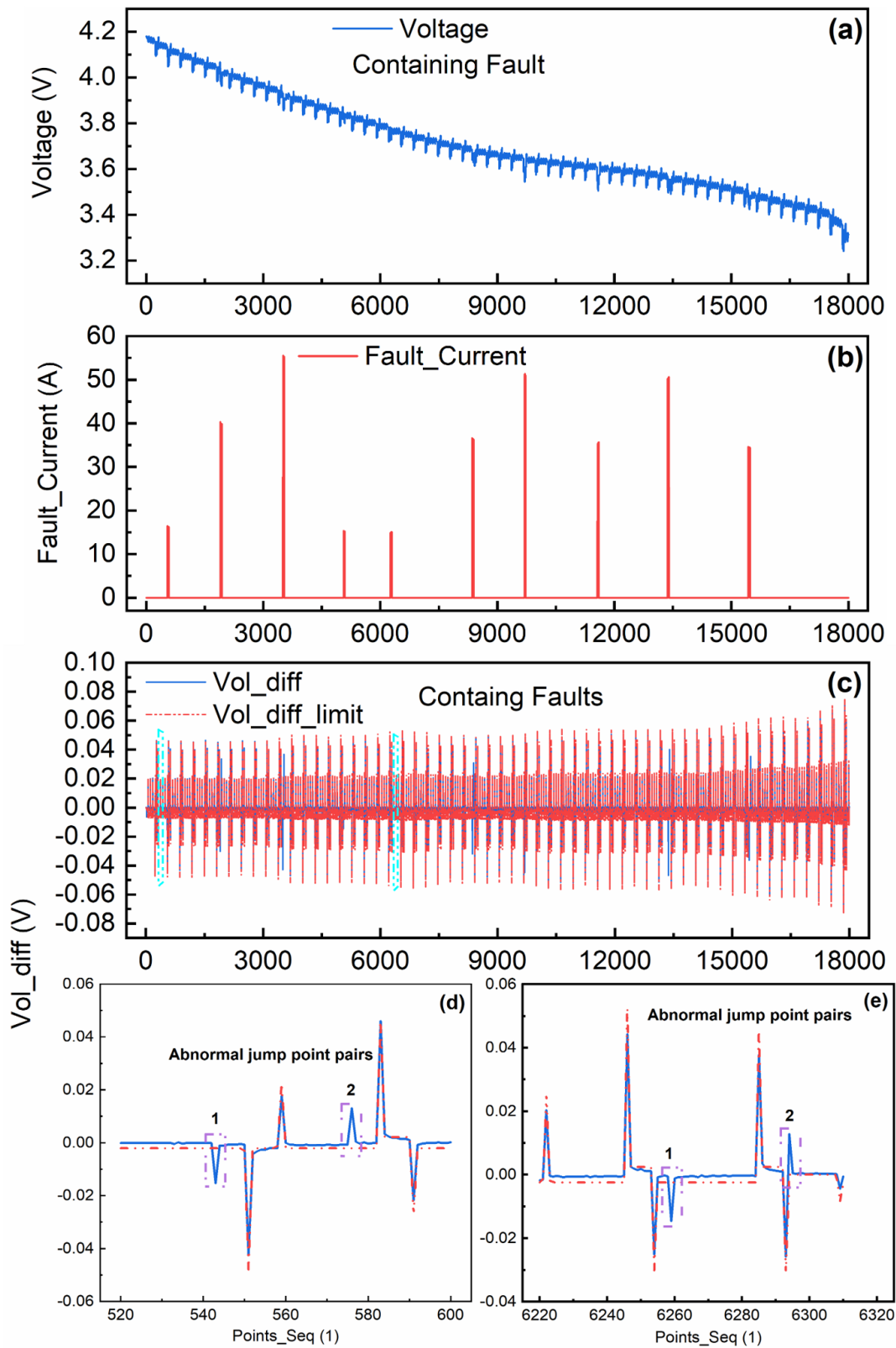

Figure 3 Electrical information of cell containing faults
(a) Voltage of battery with 10 fault intervals.
(b) Fault current pulse at various intervals.
(c) Voltage differential and envelope.
(d) The enlarged part of first fault interval of (c).

(e) The enlarged part of fifth fault interval of (c).

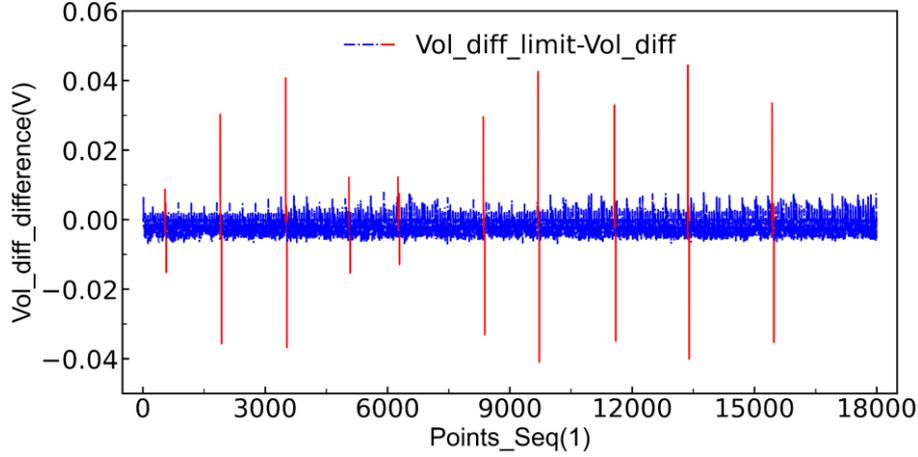

Figure 4 Difference between the voltage differential and envelope over full discharging span

## CONCLUSIONS

This paper utilizes the internal resistance and current of battery to calculate and depict the voltage differential envelope over the entire dynamic discharge period. Under normal conditions, the battery voltage differential remains within the envelope, with the maximum difference typically around 0.012 volts (when SOC < 10%). However, when a voltage drop and recovery occur due to ISC faults, the voltage differential escapes the envelope limits at the start and end points of the fault interval, showing an abnormal pattern of initial decline followed by a subsequent peak. This behavior is consistent with the characteristics of ISC faults. The proposed method was validated using ten fault intervals of varying severity. The results demonstrate that these faults can be easily identified, confirming the effectiveness of the proposed approach. Due to space constraints, this paper does not compare the method with other signal processing, information entropy theory, and deep learning-based methods. However, it is clear that the proposed method leverages the battery's intrinsic physical characteristics, providing advantages such as fast response, accurate fault localization, low computational load, and robust performance. Additionally, the algorithm is effective regardless of whether it is applied at the individual cell scale or under complex charging and discharging conditions. The algorithm fully meets the conditions for practical application, offering significant benefits for enhancing the safe usage of lithium batteries.

## METHODS

The algorithm presented in this paper involves the processing of current and voltage, utilizing the battery's internal resistance and SOC information when calculating and depicting the envelope curve. The internal resistance is obtained through pulse charging-discharging tests at different C-rates, with the maximum value being selected. The experimental details, including the specific resistance values at different SOC points, are provided in the supplementary materials. Here, only the calculation details are presented.

$$SOC_k = SOC_{k-1} + (I_k + I_{IS,k})/T/Q_c \qquad (Equation\ 1)$$
$$R_{0,k} = func(SOC_k) \qquad (Equation\ 2)$$

Where the $I$ and $I_{IS}$ are the current sampling value and fault current, T denotes sampling interval and $Q_c$ is battery capacity, $R_0$ is the internal resistance obtained by interpolation function in this study.

The processing of current differential and voltage differential is as follows:

$$\begin{cases} I_{rate,k} = round(I_k/0.05C) \\ I_{diff,k} = I_{rate,k+1} - I_{rate,k+1} \\ U_{diff,k} = U_{k+1} - U_k \end{cases} \quad (Equation\ 3)$$

Where $I_{rate}$ denote the value after round to nearest integer of the ratio of current sampling value to 0.05C (2A, battery normal capacity 40Ah), $I_{diff}$ is the current differential value and the $U_{diff}$ is the voltage differential.

The calculation of the voltage differential envelope is as follows:

If $I_{diff} < 0$

$$U_{env,k} = R_{0,k} \times (I_{diff,k} - 1) \times 0.05C \quad (Equation\ 4)$$

If $I_{diff} > 0$

$$U_{env,k} = R_{0,k} \times (I_{diff,k} + 1) \times 0.05C \quad (Equation\ 5)$$

If $I_{diff} = 0$

$$U_{env,k} = 0 \quad (Equation\ 6)$$

Here, when there is a sudden change in current, we add 1 to its variation to offset the effects of rounding and noise.

**AUTHOR CONTRIBUTIONS**

Conceptualization, C.Z.Z.; methodology, C.Z.Z.

**REFERENCES**

1. Dong, G., Chen, Z., and Wei, J. (2019). Sequential Monte Carlo Filter for State-of-Charge Estimation of Lithium-Ion Batteries Based on Auto Regressive Exogenous Model. IEEE Transactions on Industrial Electronics *66*, 8533-8544. 10.1109/tie.2018.2890499.

2. Zhang, C., Zhao, H., Wang, L., Liao, C., and Wang, L. (2024). A comparative study on state-of-charge estimation for lithium-rich manganese-based battery based on Bayesian filtering and machine learning methods. Energy *306*. 10.1016/j.energy.2024.132349.

3. Feng, X., Ren, D., He, X., and Ouyang, M. (2020). Mitigating Thermal Runaway of Lithium-Ion Batteries. Joule *4*, 743-770. 10.1016/j.joule.2020.02.010.

4. Liu, X., Ren, D., Hsu, H., Feng, X., Xu, G.-L., Zhuang, M., Gao, H., Lu, L., Han, X., Chu, Z., et al. (2018). Thermal Runaway of Lithium-Ion Batteries without Internal Short Circuit. Joule *2*, 2047-2064. 10.1016/j.joule.2018.06.015.

5. Feng, X., Ouyang, M., Liu, X., Lu, L., Xia, Y., and He, X. (2018). Thermal runaway mechanism of lithium ion battery for electric vehicles: A review. Energy Storage Materials *10*, 246-267. 10.1016/j.ensm.2017.05.013.

6. Hu, D., Huang, S., Wen, Z., Gu, X., and Lu, J. (2024). A review on thermal runaway warning technology for lithium-ion batteries. Renewable and Sustainable Energy Reviews *206*. 10.1016/j.rser.2024.114882.

7. Gao, W., Zheng, Y., Ouyang, M., Li, J., Lai, X., and Hu, X. (2019). Micro-Short-Circuit Diagnosis for Series-Connected Lithium-Ion Battery Packs Using Mean-Difference Model. IEEE Transactions on Industrial Electronics *66*, 2132-2142. 10.1109/tie.2018.2838109.

8. Zheng, Y., Lu, Y., Gao, W., Han, X., Feng, X., and Ouyang, M. (2021). Micro-Short-Circuit Cell Fault Identification Method for Lithium-Ion Battery Packs Based on Mutual Information. IEEE Transactions on Industrial Electronics *68*, 4373-4381. 10.1109/tie.2020.2984441.

9. Lai, X., Jin, C., Yi, W., Han, X., Feng, X., Zheng, Y., and Ouyang, M. (2021). Mechanism, modeling, detection, and prevention of the internal short circuit in lithium-ion batteries: Recent advances and perspectives. Energy Storage Materials *35*, 470-499. 10.1016/j.ensm.2020.11.026.

10. Zhang, G., Wei, X., Tang, X., Zhu, J., Chen, S., Dai, H., (2021). Internal short circuit mechanisms, experimental

approaches and detection methods of lithium-ion batteries for electric vehicles: A review. Renewable and Sustainable Energy Reviews *141*. 10.1016/j.rser.2021.110790.

11. Rosso, M., Brissot, C., Teyssot, A., Dollé, M., Sannier, L., Tarascon, J.-M., Bouchet, R., and Lascaud, S. (2006). Dendrite short-circuit and fuse effect on Li/polymer/Li cells. Electrochimica Acta *51*, 5334-5340. 10.1016/j.electacta.2006.02.004.
12. Orendorff, C.J., Roth, E.P., and Nagasubramanian, G. (2011). Experimental triggers for internal short circuits in lithium-ion cells. Journal of Power Sources *196*, 6554-6558. 10.1016/j.jpowsour.2011.03.035.